\documentclass{article}

\usepackage[nonatbib, preprint]{neurips_2020}

\usepackage[utf8]{inputenc} 
\usepackage[T1]{fontenc}    
\usepackage{hyperref}       
\usepackage{url}            
\usepackage{booktabs}       
\usepackage{amsfonts}       
\usepackage{nicefrac}       
\usepackage{microtype}      
\usepackage{graphicx}
\usepackage{wrapfig}

\usepackage{floatrow}
\newfloatcommand{capbtabbox}{table}[][\FBwidth]

\usepackage{amsmath,amsthm,amssymb,mathrsfs,mathabx}
\usepackage[nameinlink]{cleveref}

\title{Enhancing SAT solvers with glue variable predictions}

\author{%
  Jesse Michael Han\thanks{Work done while part of the Automated Reasoning Group at Amazon Web Services.}
  \\
  Department of Mathematics\\
  University of Pittsburgh\\
  Pittsburgh, PA 15213 \\
  \texttt{jmh288@pitt.edu} \\
}
\newcommand{\cadical}{{\sc CaDiCaL}}
\newcommand{\wh}{\widehat}

\graphicspath{ {./graphics/} }

\begin{document}

\maketitle

\begin{abstract}
  Modern SAT solvers routinely operate at scales that make it impractical to query a neural network for every branching decision. NeuroCore, proposed by \cite{DBLP:conf/sat/SelsamB19}, offered a proof-of-concept that neural networks can still accelerate SAT solvers by only periodically refocusing a score-based branching heuristic. However, that work suffered from several limitations: their modified solvers require GPU acceleration, further ablations showed that they were no better than a random baseline on the SATCOMP 2018 benchmark, and their training target of unsat cores required an expensive data pipeline which only labels relatively easy unsatisfiable problems. We address all these limitations, using a simpler network architecture allowing CPU inference for even large industrial problems with millions of clauses, and training instead to predict {\em glue variables}---a target for which it is easier to generate labelled data, and which can also be formulated as a reinforcement learning task. We demonstrate the effectiveness of our approach by modifying the state-of-the-art SAT solver {\sc CaDiCaL}, improving its performance on SATCOMP 2018 and SATRACE 2019 with supervised learning and its performance on a dataset of SHA-1 preimage attacks with reinforcement learning.
\end{abstract}

\section{Introduction}
Branching heuristics for search procedures in automated theorem provers are an attractive target for deep learning methods, and have been the focus of recent work ranging from higher-order theorem proving \cite{DBLP:conf/icml/BansalLRSW19} and first-order theorem proving \cite{DBLP:conf/lpar/LoosISK17}, to QBF solving \cite{DBLP:conf/iclr/LedermanRSL20, DBLP:journals/corr/abs-1904-12084} and SAT solving \cite{DBLP:conf/sat/SelsamB19, neural-heuristics-for-sat, kurin2019improving, DBLP:conf/nips/YolcuP19}.
Branching heuristics for SAT solving are a particularly challenging target for applying deep learning, as modern SAT solvers are heavily optimized and routinely operate at scales (tens of thousands of decisions per second, problems with millions of variables and clauses) that make it impractical to query a neural network for every branching decision.

There are many design decisions whose trade-offs which must be carefully balanced in order to efficiently integrate machine learning into the branching heuristics of a modern SAT solver. These include whether or not to condition on the solver state and history, whether the model is trained on- or offline, how to integrate predictions into the solver, and the trade-off between model capacity and inference time. Perhaps most important is whether the model is conditioned on the global problem state. The CDCL algorithm, as typically implemented with lazy data structures \cite{DBLP:conf/dac/MoskewiczMZZM01}, performs essentially \emph{local} probing of the instance being solved: the solver only tracks the direct consequences of unit propagation of its assignment stack, and is almost never aware of the global state of the problem, i.e. 
how the problem actually simplifies under the current assignment. Doing so would require traversing every clause in the clause database, an operation which can take multiple seconds on large problems. Thus, any globally-informed heuristic already carries an enormous upfront cost---the execution of the solver must be halted and all clause pointers dereferenced. This cost must be amortized by the quality of the heuristic's decisions, incentivizing higher-capacity models such as neural networks.

To date, however, the most successful applications of machine learning to branching heuristics in SAT solving have appeared in the Maple family of solvers \cite{DBLP:phd/basesearch/Liang18a, DBLP:conf/sat/LiangGPC16}, involving low-capacity, non-deep models for which inference is instantaneous, trained online during the execution of the solver, and queried for every branching decision, conditioned only on local information obtained from conflict analysis. Maple solvers have either won or placed highly in the annual international SAT competition \cite{satcomp-website} since 2016.

In contrast, much of the existing work in applying deep learning to SAT and QBF solving uses globally-conditioned graph neural networks \cite{DBLP:journals/tnn/ScarselliGTHM09} with much more expensive inference times to either solve trivially small problems end-to-end \cite{DBLP:conf/iclr/SelsamLBLMD19, DBLP:conf/iclr/AmizadehMW19, DBLP:journals/corr/abs-1904-12084} or guide a search algorithm on every branching decision \cite{DBLP:conf/iclr/LedermanRSL20, neural-heuristics-for-sat, kurin2019improving,DBLP:conf/nips/YolcuP19} on problems with at most a few thousand clauses. It is unlikely that these methods can scale to large, real-world use-cases as represented in the SAT competitions. The most promising step in this direction is NeuroCore \cite{DBLP:conf/sat/SelsamB19}, which trained a neural network to predict unsat cores and avoids the performance overhead of querying for every branching decision by using the network's predictions to only \emph{periodically refocus} a score-based branching heuristic. However, their work had several limitations: (1) their modified solvers required GPUs for inference and were vastly more expensive to run than the CPU-only base solvers; (2) they only modified the SAT solvers Minisat and Glucose, which are no longer state of the art, and further ablations \cite{DBLP:journals/corr/abs-1903-04671} showed that their modified solvers were no better than a random baseline on the SATCOMP 2018 benchmark; and (3) in order to produce enough labelled unsat cores for training, they relied on an expensive data pipeline that only labels relatively easy unsatisfiable problems.

In our present work, we address all these issues and show that we can realize the promise of using neural networks to accelerate state-of-the-art SAT solvers with no additional hardware. First, we use a simpler network architecture, allowing CPU inference for even large industrial problems with millions of clauses. Second, instead of unsat cores, we train to predict \emph{glue variables}---those likely to occur in glue clauses, a type of conflict clause known to be extremely important to the reasoning of modern CDCL SAT solvers. Glue clauses arise frequently during search and do not require a solver to run to completion in order to generate training data. Finally, we target the state-of-the-art solver {\sc CaDiCaL}, and achieve improvements over the unmodified solver and a random baseline on SATCOMP 2018 and SATRACE 2019.\footnote{\url{https://www.github.com/jesse-michael-han/neuro-cadical/}}
We also show that glue variable prediction can be formulated in terms of reinforcement learning; we use this to learn distribution-specific heuristics and improve solver performance on a dataset of problems encoding SHA-1 preimage attacks.

\section{Background} \label{sect:background}
A propositional logic formula is a Boolean expression (i.e. using the unary negation operator \(\neg\) and the binary operators \(\land\) and \(\lor\)) of the constants \(0\) (false), \(1\) (true), and variables. A \emph{literal} is a variable or a negation of a variable. A \emph{clause} is a disjunction of literals. A formula is in \emph{conjunctive normal form} (CNF) if it is a conjunction of clauses; every formula in propositional logic is equivalent to a CNF formula \cite{tseitin1983complexity}. The satisfiability problem (SAT) for propositional logic is to find an assignment (to \(1\) or \(0\)) of all variables of a given formula (sometimes called \emph{instance}) \(\phi\) such that the formula is equivalent to \(1\) (i.e. \(\phi\) is satisfiable), or prove that no such assignment exists (i.e. \(\phi\) is unsatisfiable). The formula \(\phi\) will typically be in CNF, in which case satisfiability is equivalent to simultaneously being able to satisfy all clauses of \(\phi\). SAT is the prototypical NP-complete problem \cite{DBLP:conf/stoc/Cook71}.

The DPLL algorithm \cite{DBLP:journals/jacm/DavisP60, DBLP:journals/cacm/DavisLL62} was introduced as a complete decision procedure for propositional logic. It heavily relies on \emph{unit propagation}. Given a clause \(C = (\ell_1 \lor \dots \lor \ell_n)\), if all literals but \(\ell_n\) are set to \(0\), then \(C\) is equivalent to the \emph{unit clause} \((\ell_n)\), and the value of \(\ell_n\) is forced to \(1\). This may lead to simplifications elsewhere which produce more unit clauses; unit propagation refers to repeating this procedure of identifying unit clauses and propagating simplifications until fixpoint.

The DPLL algorithm was significantly extended by the \emph{conflict-driven clause learning} (CDCL) algorithm \cite{DBLP:conf/ictai/SilvaS96}, which is now the dominant paradigm in SAT solving. CDCL performs \emph{conflict analysis} before backtracking: upon encountering a conflict, a CDCL solver analyzes the directed acyclic graph of unit propagations leading to the conflict and derives a \emph{conflict clause} which, when added to the formula, prunes the part of the search tree which led to the conflict. Each new conflict clause is justifiable from existing clauses by a sequence of resolution steps; if the formula is unsatisfiable, then eventually the empty clause will be derived from conflict analysis, resulting in a proof of unsatisfiability. 
Unsatisfiability tends to be harder to prove than satisfiability, because all candidate assignments must be ruled out; generally, exponential lower bounds on DPLL runtime arise from families of unsatisfiable formulas \cite{cook1979relative, cook1976short, DBLP:journals/apal/BeameP96, DBLP:journals/tcs/Alekhnovich04}.

\paragraph{Glue levels and glue clauses}

A key insight of the Glucose series of solvers \cite{DBLP:journals/ijait/AudemardS18} is that the quality of a conflict clause can be approximated by its literal block distance (LBD), or \emph{glue level} \cite{DBLP:conf/ijcai/AudemardS09}, which counts the number of decision levels involved in the clause. A clause with low glue level requires fewer decisions to become unit, and will be disproportionately involved in unit propagation after being added to the clause database. Clauses with glue level \(\leq 2\) are called \emph{glue clauses}, and are so important that Glucose never removes them from its clause database while aggressively deleting high-LBD clauses; this led to massive performance gains over the existing state-of-the-art and is now standard practice.

\paragraph{Score-based branching heuristics} The Variable State-Independent Decaying Sum (VSIDS) heuristic and its more popular variant Exponential-VSIDS (EVSIDS) have been the dominant branching heuristic in CDCL SAT solvers for over a decade \cite{DBLP:conf/dac/MoskewiczMZZM01}. EVSIDS greedily selects decision variables according to an \emph{activity score} maintained for each variable, which is modified during conflict analysis: if a variable is involved in a conflict, its activity is \emph{bumped} by a fixed increment, and after every conflict, regardless of participation, every variable's activity score is multiplicatively \emph{decayed} by a factor \(0 < \rho < 1\). Thus, variables which participate frequently in conflicts will have higher activity scores, weighted towards more recent conflicts.

\section{Network architecture} \label{sect:network-architecture}

We describe our network architecture, which is similar to the graph neural network used in \cite{DBLP:conf/iclr/LedermanRSL20}. We encode a SAT problem with \(N\) variables and \(M\) clauses as the \(M \times 2 \cdot N\) sparse bipartite adjacency matrix \(G\) of its \emph{clause-literal graph}, which has a node for every clause and literal, and an edge between a literal \(\ell\) and a clause \(C\) iff \(\ell\) occurs in \(C\). The adjacency matrix \(G\) is the input to our network, which is parametrized by the following learnable parameters:

\begin{itemize}
\item An initial literal embedding \(\mathbf{l}_{\text{init}}\)
\item An \(n_C\)-layer feedforward network \(C_{\text{update}} : \mathbb{R}^{2\cdot\delta_L} \to \mathbb{R}^{\delta_C}\)
\item An \(n_L\)-layer feedforward network \(L_{\text{update}} : \mathbb{R}^{\delta_C} \to \mathbb{R}^{\delta_L}\)
\item An \(n_P\)-layer feedforward network \(V_{\text{policy}} : \mathbb{R}^{2 \cdot \delta_L} \to \mathbb{R}^1\)
\item A layer normalization \cite{DBLP:journals/corr/BaKH16} \(\mathrm{LayerNorm} : \mathbb{R}^{\delta_L} \to \mathbb{R}^{\delta_L}\).
\end{itemize}

The network computes forward as follows. For every literal \(\ell\), we initialize an embedding \(\mathbf{l} = \mathbf{l}_{\text{init}}\). Let \(\overline{\ell}\) denote the negation of \(\ell\), and let \(\overline{\mathbf{l}}\) denote the embedding of \(\overline{\ell}\). Let \(\mathbf{L}\) denote the \(2 \cdot N_{\text{var}} \times \delta_L\) array of all literal embeddings, and let \(\overline{\mathbf{L}}\) denote the operation of interchanging each row \(\mathbf{l}\) of \(\mathbf{L}\) with \(\overline{\mathbf{l}}\).

We compute the clause and literal embeddings as follows. In what follows, function application notation denotes row-wise application. For up to \(\tau\) iterations, we perform the following updates:
\begin{align*}
  \mathbf{C} &\leftarrow C_{\text{update}}(G \cdot \text{Concat}(\mathbf{L}, \overline{\mathbf{L}})) \\
  \mathbf{C} & \leftarrow \dfrac{\mathbf{C} - \mathbb{E}[\mathbf{C}]}{\sqrt{\text{Var}(\mathbf{C}) + \varepsilon}} \\
  \mathbf{L} & \leftarrow L_{\text{update}}(G^T \cdot \mathbf{C})  + 0.1 \cdot \mathbf{L}\\
  \mathbf{L} &\leftarrow \mathrm{LayerNorm}(\mathbf{L}).
\end{align*}

Finally, we obtain a probability distribution \(\wh{\pi}\) over variables by applying \[\wh{\pi} \leftarrow \mathrm{Softmax} \left( V_{\text{policy}} \left(\text{Concat}(\mathbf{L}, \overline{\mathbf{L}})\right)  \right).\]

Above, \(\delta_C\) is the dimension of the clause embeddings, and \(\delta_L\) is the dimension of the literal embeddings; the number of iterations \(\tau\), the number of layers \(n_L, n_C, n_P\) in the feedfoward networks, and \(\delta_L, \delta_C\) are hyperparameters. We use LeakyReLU nonlinearities for the hidden layers of the feedforward networks, and during training, we use a dropout fraction of 0.15 throughout. For reinforcement learning, we attach a value head \(V_{\text{value}}\) which is identical to \(V_{\text{policy}}\), and obtain a value estimate \(\wh{v}\) with
\[\wh{v} \leftarrow \mathrm{Sigmoid}\left(\mathrm{Mean}\left(V_{\text{value}}(\mathrm{Concat}(\mathbf{L}, \overline{\mathbf{L}}))\right)\right).\]

The main differences between our architecture and the one used in \cite{DBLP:conf/iclr/LedermanRSL20} are the normalization of clause and literal embeddings, the residual layer during the literal update step, the absence of clause and literal features extracted from the solver state, and our choice of hyperparameters, which are tuned for more expensive and less frequent queries instead of querying for every decision. Importantly, in contrast to NeuroSAT-style architectures \cite{DBLP:conf/iclr/SelsamLBLMD19, neural-heuristics-for-sat} which update \(\mathbf{L}\) using \(G^T \cdot \mathbf{C}\) and \(\overline{\mathbf{L}}\), this architecture updates \(\mathbf{C}\) using \(G \cdot \mathbf{L}\) and \(G \cdot \overline{\mathbf{L}}\). When the number of iterations increases beyond \(\tau = 1\), this ensures that every clause embedding is updated partly according to the embeddings of its possible resolvents, i.e. the message-aggregation step for clause embeddings indirectly incorporates the structure of the resolution graph of the formula.

\section{Data generation and training} \label{sect:pipeline}
\subsection{Supervised learning of glue variable prediction} 
\label{subsect:supervised-learning}
We modify the solver {\sc CaDiCaL} \Cref{sect:cadical} to halt after 180 seconds and traverse the clause database, accumulating counters for the number of times each variable appears in a glue clause. These glue counts, along with the sparse clause-literal adjacency matrix of the original formula, form a single datapoint. We perform this procedure for all \(750\) main track problems in SATCOMP 2016 and SATCOMP 2017. To ensure uniformity in training data, if necessary, we split a problem into subproblems by randomly assigning variables until the resulting subproblems each have \(\leq 150000\) clauses. We synthetically augment our dataset by periodically dumping the entire formula plus learned clauses every 100000 conflicts, then running the data-generation procedure again.

We generated a training set of approximately \(50000\) datapoints. During training, we softmax the glue counts to obtain a probability distribution \(\pi\) and train to minimize the KL divergence between \(\pi\) and the probability distribution \(\wh{\pi}\) emitted by the network. We used the hyperparameters
\(\delta_L = 16, \delta_C = 64, \tau = 2, n_L = 2, n_c = 2, n_P = 3\), choosing relatively small values in anticipation of the large industrial problems in the evaluation set.
   We trained for 3 epochs with averaged stochastic gradient descent \cite{10.1137/0330046} with learning rate 1e-3, using RaySGD \cite{DBLP:conf/osdi/MoritzNWTLLEYPJ18} and data-distributed Pytorch \cite{DBLP:conf/nips/PaszkeGMLBCKLGA19} on 32 GPUs in under an hour.

\subsection{Reinforcement learning of glue level minimization} 
\label{subsect:reinforcement-learning}

Motivated by recent work \cite{DBLP:conf/iclr/LedermanRSL20, kurin2019improving, DBLP:conf/nips/YolcuP19} showing that reinforcement learning techniques can learn effective distribution-specific branching heuristics from only dozens or hundreds of training problems, we frame glue variable prediction for a formula \(\phi\) as an episodic reinforcement learning task on a finite Markov decision process (MDP), represented by the data \(\mathcal{S}_{\phi}, \mathcal{A}, \mathcal{T}, \mathcal{R}\) defined as follows:

\begin{itemize}
\item The collection of possible states \(\mathcal{S}_{\phi}\) comprises all subformulas of \(\phi\), i.e CNFs obtained by simplifying and applying unit propagation to \(\phi\) with respect to partial assignments. The environment enters a terminal state when all variables have been assigned. There are two terminal states, corresponding to whether the formula is satisfied or unsatisfied.
\item \(\mathcal{A}\) assigns to each \(s \in \mathcal{S}_{\phi}\) the collection of valid actions; these are just the variables which have not yet been assigned.
\item \(\mathcal{T} : \Pi_{s \in \mathcal{S}_{\phi}} \mathcal{A}(s) \to \mathcal{S}_{\phi}\) is a stochastic transition function. Once a variable has been selected for assignment, the environment assigns it either \(0\) or \(1\) with uniform probability and simplifies the formula with unit propagation to obtain the next state.
\item \(\mathcal{R}\) is the reward function on states. For non-terminal states, we always assign the small negative reward \(-1/n\), where \(n\) is the number of variables in \(\phi\). Upon reaching a terminal state, we assign \(0\) reward if the formula has been satisfied, and otherwise assign \(1/g^2\), where \(g\) is the glue level of the conflict clause learned from conflict analysis.
\end{itemize}

Note that in contrast to previous work in this vein \cite{DBLP:conf/iclr/LedermanRSL20, kurin2019improving, DBLP:conf/nips/YolcuP19}, we use domain-specific knowledge to replace a sparse terminal reward (completely solving \(\phi\)) with a proxy reward (minimizing glue level of learned clauses) that allows us to ignore backtracking and treat every path through the DPLL search tree as a separate episode.

We convert {\sc CaDiCaL} into a reinforcement learning environment which implements the dynamics outlined above. Upon receiving an action (a variable to assign), the environment sets it to a random polarity, performs unit propagation, and returns an observation in the form of a sparse clause-literal adjacency matrix, constructed as described in \Cref{sect:cadical}, along with the reward. Note that in keeping with the formal definition, we avoid non-stationarity in the environment by discarding any learned conflict clauses when resetting from a terminal state.

\paragraph{The \texttt{sha-1} dataset} We evaluate our reinforcement learning pipeline using a dataset of \(250\) formulas encoding SHA-1 preimage attacks, with an 80-20 train-test split. We generate the dataset using the tool {\sc CGen} \cite{vsklad1, cgen}. The \texttt{sha-1} dataset is of similar difficulty to a collection of {\sc CGen}-generated problems submitted to SATRACE 2019 \cite{vsklad2}. 
We generate a new six-character alphanumeric message string and randomly set the number of message variables between \(70\) and \(90\) for every instance in our dataset, leaving all other arguments to {\sc CGen} the same as the submitted benchmark.

\paragraph{Training} Since all the problems in the \texttt{sha-1} dataset are around the same size and relatively small (\(\approx 3600\) variables and \(\approx 15000\) clauses), we use the more expensive hyperparameters \(\delta_L = 32, \delta_C = 64, \tau = 4, n_L = 3, n_c = 3, n_P = 4\). We train our network (\Cref{sect:network-architecture}) using synchronous multi-agent REINFORCE with a jointly-learned value function baseline. For each batch, each member of a pool of 128 workers equipped with the latest policy independently samples a formula \(\phi\) from the training set and generates \(b\) episodes, which are then processed by a GPU learner using the Adam optimizer with constant learning rate 1e-4. Besides standard optimizations like advantage normalization and gradient clipping, we use importance sampling to correct for policy lag across multiple gradient steps.

\section{Solver modifications} \label{sect:cadical}

As with NeuroCore \cite{DBLP:conf/sat/SelsamB19}, we avoid the performance overhead of querying our network for every branching decision by only \emph{periodically refocusing} the EVSIDS branching heuristic with our trained networks (henceforth called NeuroGlue). We modify {\sc CaDiCaL} \cite{biere2017cadical, biere2018cadical, biere2019cadical}, a state-of-the-art CDCL SAT solver which solved the most instances in SAT Race 2019 and won the unsatisfiable track in SATCOMP 2018. We use the version submitted to SAT Race 2019, which incorporates an EVSIDS branching heuristic during certain phases of search \cite{DBLP:conf/sat/Oh15}
. In order to more accurately measure the impact of periodic refocusing, which only updates EVSIDS scores, we run {\sc CaDiCaL} in its \texttt{---sat} configuration, which specializes for satisfiable instances and exclusively relies on EVSIDS for branching. With this configuration, {\sc CaDiCaL} is still state-of-the-art on satisfiable instances, winning the satisfiable track of SAT Race 2019 (and beating Glucose 4.1 on the main track regardless).

\paragraph{Implementation of periodic refocusing}

We implement periodic refocusing as an inprocessing routine in \cadical which fires immediately before the next decision variable is selected, guarded by a scheduling heuristic (\Cref{cadical:refocus-schedule}).

When refocusing, we construct a sparse clause-literal adjacency matrix \(G\) by traversing all the original, then learned, clauses, simplifying with respect to the current assignment and compacting assigned variables. Like NeuroCore, we stop when the number of edges in the clause-literal graph exceeds a predetermined threshold of 10e6; if the original clauses do not meet this cutoff, then we do not query at all. \(G\) is the input to our model, which is invoked via TorchScript \cite{torchscript} and runs in the same thread as the solver. The returned logits are multiplied by a temperature parameter \(\tau = 4.0\), then softmaxed to produce a probability distribution \(\wh{\pi}\) over variables. These probabilities are then rescaled by the number of variables and a fixed constant \(\kappa = 1e4\) before replacing the existing EVSIDS scores.

\paragraph{Refocusing schedule}
\label{cadical:refocus-schedule}
Unlike \cite{DBLP:conf/sat/SelsamB19}, in which periodic refocusing is scheduled according to wall-clock intervals, we use a conflict schedule instead, performing the \(N^{\operatorname{th}}\) refocus after \[\operatorname{min}(50000 + 1000 \cdot (N-1)^2, 250000)\]
conflicts have occurred. Immediately after starting the solver, we also allot a fixed 15-second ``warm-up'' period during which no refocusing occurs.

Glucose introduced a dynamic restart strategy \cite{DBLP:journals/ijait/AudemardS18} based on an exponential moving average (EMA) of glue levels: if the EMA is significantly worse than the global average of glue levels, a restart is triggered. {\sc CaDiCaL} uses a similar strategy to schedule restarts, replacing the global average with a slower-moving EMA. As a final optimization, we incorporate this statistic into the refocusing schedule: a refocus is triggered if and only if the conflict schedule has been satisfied and the fast glue level EMA is \(10\%\) higher than the slow glue level EMA. In practice, this occurs quite often, so the overall frequency of refocusing is unaffected.

\section{Experiments} \label{sect:experiments}
We evaluate three versions of \cadical: (1) \texttt{neuro-cadical}, which performs periodic refocusing using NeuroGlue, (2) \texttt{vanilla-cadical}, the unmodified baseline, and (3) \texttt{random-cadical}, a random baseline with identical logic to \texttt{neuro-cadical}, except that during refocusing, the logits obtained from NeuroGlue are replaced by logits uniformly and independently sampled from \([0,1]\).

All evaluation runs were done in parallel on a cluster of ten \texttt{r5d.24xlarge} AWS EC2 instances with 48 cores and 768GB RAM each, and no hyperthreading.

\begin{figure}
    \centering
    \begin{minipage}{0.49\textwidth}
        \centering
        \includegraphics[width=1.02\textwidth]{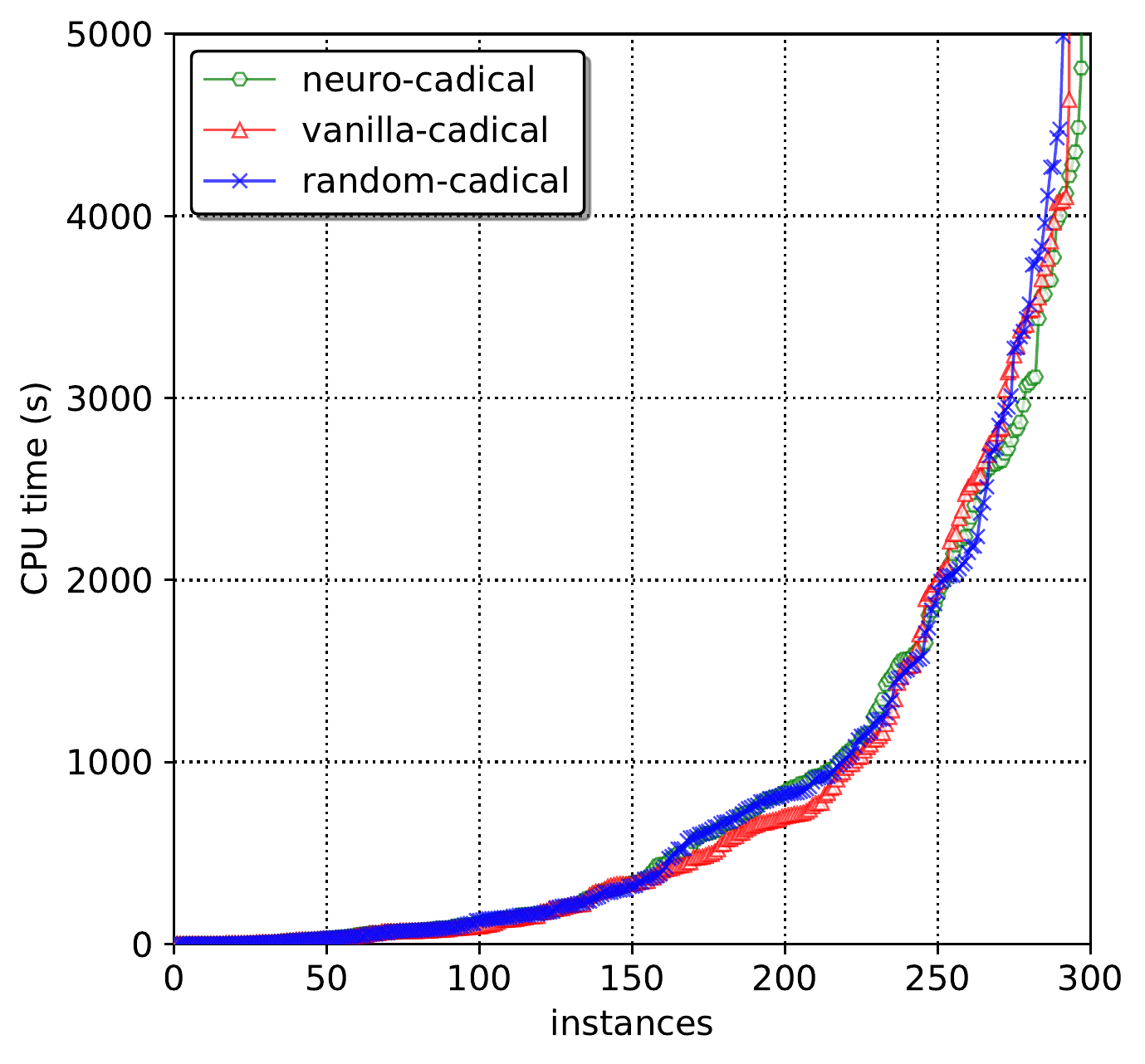}
        \label{fig:sc18-rtime}
    \end{minipage}\hfill
    \begin{minipage}{0.49\textwidth}
        \centering
  \includegraphics[width=1.01\textwidth]{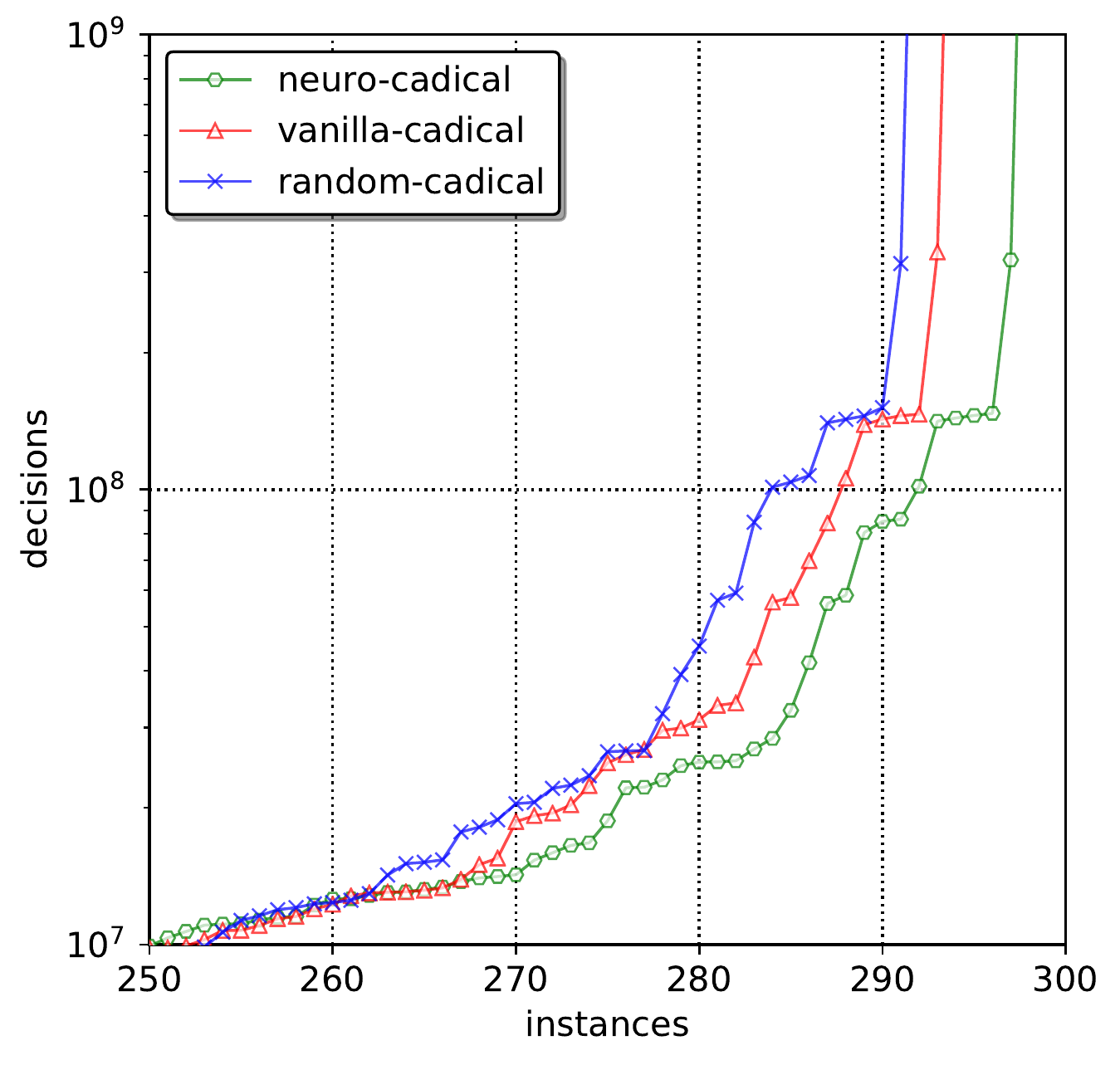}
        \label{fig:sc18-decisions}
      \end{minipage}
    \centering
    \begin{minipage}{0.49\textwidth}
        \centering
        \includegraphics[width=1.0\textwidth]{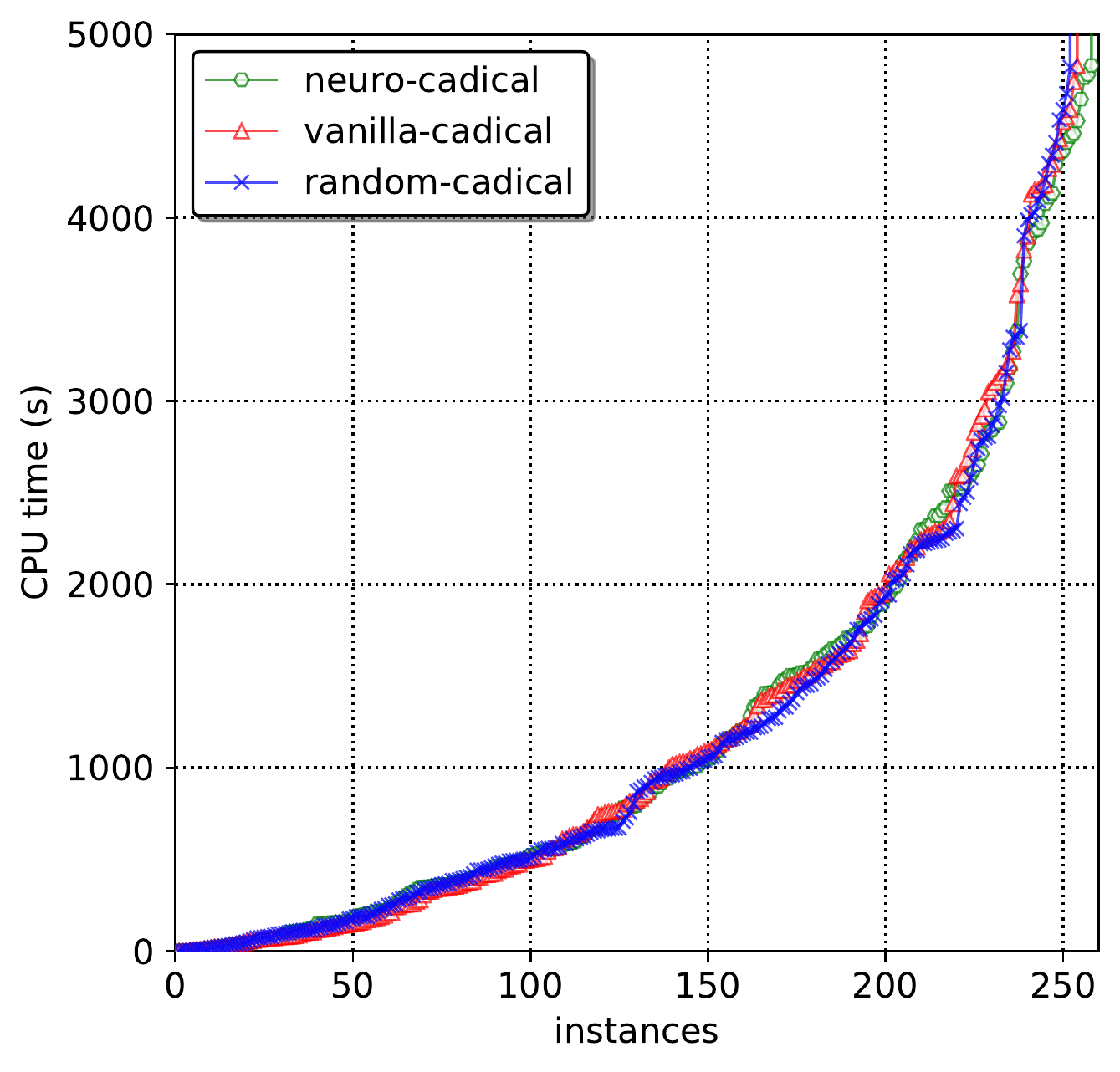}
        \label{fig:sr19-rtime}
    \end{minipage}\hfill
    \begin{minipage}{0.49\textwidth}
        \centering
  \includegraphics[width=1.01\textwidth]{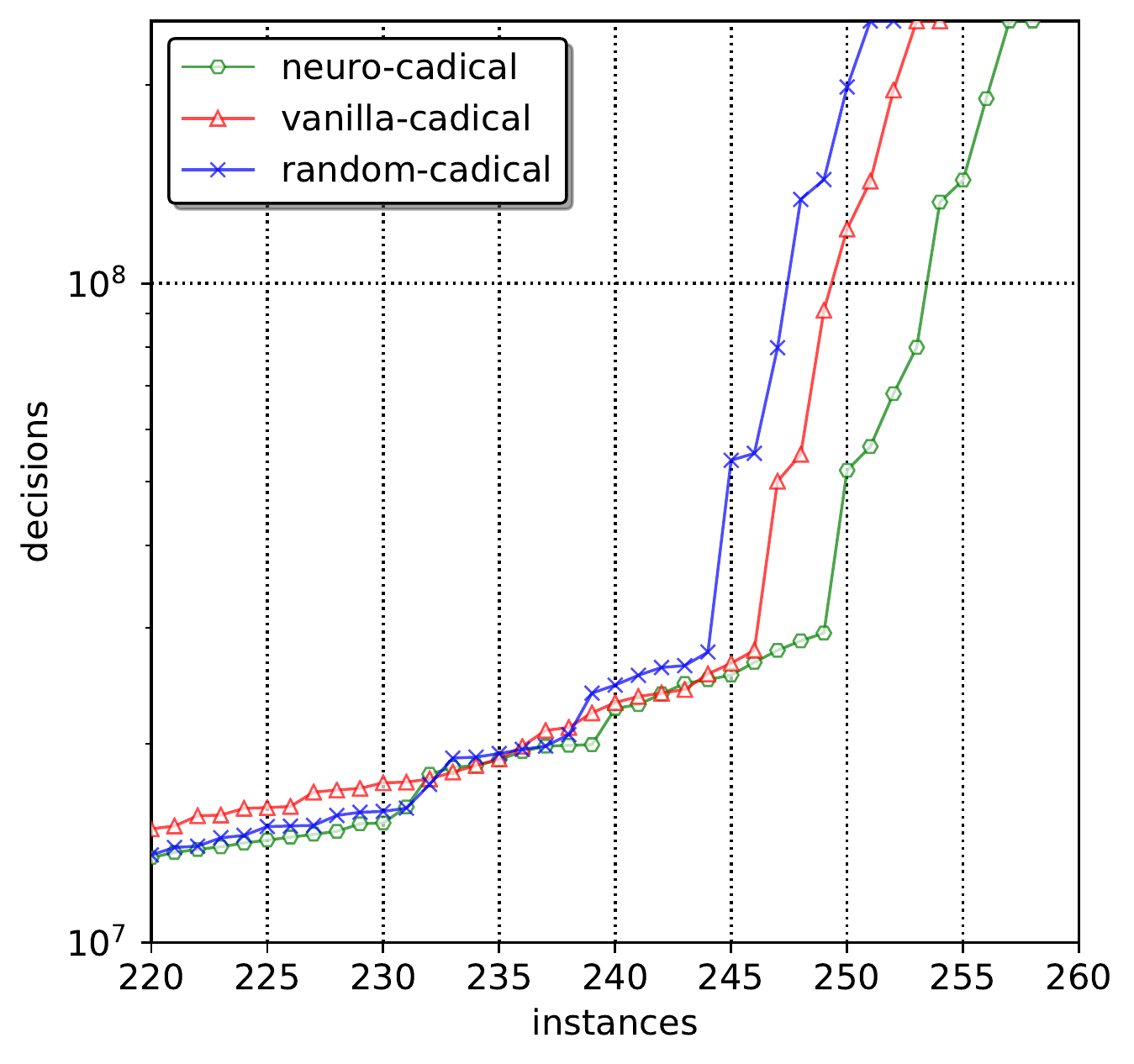}
        \label{fig:sr19-decisions}
      \end{minipage}
      \caption{
        \textbf{Left:} Runtime cactus plots of all three variants on {\sc CaDiCaL} on SATCOMP 2018 (top) and SATRACE 2019 (bottom). In both cases, most of \texttt{neuro-cadical}'s lead is accumulated from more difficult problems towards the end. \textbf{Right:} Decision cactus plots on SATCOMP 2018 with 175M decision limit, starting at the 250 problem cutoff (top) and on SATRACE 2019 with 250M decision limit, starting at the 220 problem cutoff (bottom). In both cases, the improved runtime of \texttt{neuro-cadical} is reflected in its superior decision efficiency over both baselines.
      }
      \label{fig:sc-cactus-plots}
      \end{figure}

\subsection{SATCOMP 2018 and SATRACE 2019} \label{subsect:sc-evals}
In keeping with the rules of recent SAT competitions \cite{satcomp-website}, each solver process runs with a 5000 second timeout, and our primary metric is the PAR-2 score, defined as
the sum of runtimes for all solved instances plus \(2 \cdot \text{timeout} \cdot (\text{\# of unsolved instances})\), so that
a lower PAR-2 score is better. For conciseness, we divide all PAR-2 scores by the total number of instances. We also measure the \emph{global learning rate} (GLR) i.e. ratio of conflicts to decisions, and \emph{average glue level} for each problem, which have been empirically shown to be correlated with the quality of a branching heuristic \cite{DBLP:conf/ijcai/LiangKPCG18}.

As a measure against noise due to resource contention or unlucky random seeds for either \texttt{random-cadical} or randomized heuristics in the base solver, we perform 16 evaluation runs with distinct random seeds on SATCOMP 2018 and SATRACE 2019, averaging the results for each of the 400 instances in each benchmark. We consider solver \(S\) to have solved \(\phi\) if in any of the evaluation runs, \(S\) solved \(\phi\), and we only average the successful runtimes of \(S\) on \(\phi\). This resembles the construction of a virtual best solver \cite{DBLP:conf/issac/BrainDG17}, except we compare a solver against only itself across evaluation runs instead of against all other solvers, and take the average of successful runtimes instead of the minimum. We additionally calculate PAR-2 scores for satisfiable (resp. unsatisfiable) instances by restricting the score calculation to instances which were found to be satisfiable (resp. unsatisfiable) by any of the three solvers.

Our results are shown in \Cref{tab:sc18} and \Cref{tab:sr19}. On SATCOMP 2018, \texttt{neuro-cadical} achieves a \(1.67\%\) better score than \texttt{vanilla-cadical} and a \(2.98\%\) better score than \texttt{random-cadical}. On SATRACE 2019, \texttt{neuro-cadical} achieves a 1.38\% better score than \texttt{vanilla-cadical} and a 1.68\% better score than \texttt{random-cadical}. To put this in perspective, the margin in PAR-2 scores between first and second place in the three most recent SAT competitions averages \(1.15\%\). \Cref{fig:sc-cactus-plots} displays runtime and decision cactus plots on both datasets. In both cases, most of \texttt{neuro-cadical}'s lead is accumulated from more difficult problems towards the end of the plots, and it requires fewer decisions to solve more of these problems. \Cref{tab:sc18-comp} and \Cref{tab:sr19-comp} display the proportion of each benchmark that each solver attained better GLR and better average glue level. On both datasets, \texttt{neuro-cadical} has better GLR on more problems than both baselines.

\begin{table}
  \RawFloats
    \centering
    \begin{minipage}{0.50\textwidth}
  \caption{PAR-2 scores on SATCOMP 2018.} \label{tab:sc18}
  \centering
  \hspace{9mm}
  \begin{tabular}{llll}
    \toprule
    Solver     & overall     & sat & unsat \\
    \midrule
    \texttt{neuro}   & 3194.81 & 969.68   & 1194.35 \\
    \texttt{vanilla} & 3249.99 & 1090.52  & 1172.87 \\
    \texttt{random}  & 3293.62 & 1033.93  & 1452.72 \\
    \bottomrule
  \end{tabular}

    \end{minipage}\hfill
    \begin{minipage}{0.50\textwidth}
  \caption{PAR-2 scores on SATRACE 2019.} \label{tab:sr19}
  \centering
  \hspace{9mm}
  \begin{tabular}{llll}
    \toprule
    Solver     & overall     & sat & unsat \\
    \midrule
    \texttt{neuro}   & 4344.87 & 1039.68   & 1817.05 \\
    \texttt{vanilla} & 4405.69 & 1134.12  & 1908.87 \\
    \texttt{random}  & 4419.33 & 1155.15  & 1929.76 \\
    \bottomrule
  \end{tabular}

      \end{minipage}
    \end{table}

\begin{table}
  \RawFloats
    \centering
    \begin{minipage}{0.48\textwidth}
  \caption{Percent of problems on SATCOMP 2018 with better GLR/average glue level.} \label{tab:sc18-comp}
  \centering
  \hspace{9mm}
  \begin{tabular}{lll}
    \toprule
                     & higher GLR     & lower avg glue  \\
    \midrule
    \texttt{neuro}   & 66.5\% & 54.9\%   \\
    \texttt{vanilla} & 33.5\% & 45.1\%  \\
    \midrule
    \texttt{neuro}   & 56.2\% & 54.9\%   \\
    \texttt{random}  & 43.8\% & 45.1\%  \\
    \midrule
    \texttt{vanilla} & 32.5\% & 53.0\%   \\
    \texttt{random}  & 67.5\% & 47.0\%  \\
    \bottomrule
  \end{tabular}

    \end{minipage}\hfill
    \begin{minipage}{0.48\textwidth}
\caption{Percent of problems on SATRACE 2019 with better GLR/average glue level.} \label{tab:sr19-comp}
  \centering
  \hspace{9mm}
  \begin{tabular}{lll}
    \toprule
                     & higher GLR     & lower avg glue  \\
    \midrule
    \texttt{neuro}   & 54.2\% & 48.8\%   \\
    \texttt{vanilla} & 45.8\% & 51.2\%  \\
    \midrule
    \texttt{neuro}   & 52.5\% & 45.9\%   \\
    \texttt{random}  & 47.5\% & 54.1\%  \\
    \midrule
    \texttt{vanilla} & 46.2\% & 50.4\%   \\
    \texttt{random}  & 53.8\% & 49.6\%  \\
    \bottomrule
  \end{tabular}

      \end{minipage}
\end{table}

\begin{wraptable}{r}{200pt}
  \caption{Percent of problems on the \texttt{sha-1} test set with better GLR/average glue level.} \label{tab:sha250-comp}    
  \centering
  \hspace{9mm}

  \begin{tabular}{lll}
    \toprule
                     & higher GLR     & lower avg glue  \\
    \midrule
    \texttt{neuro}   & 60.0\% & 50.0\%   \\
    \texttt{vanilla} & 40.0\% & 50.0\%  \\
    \midrule
    \texttt{neuro}   & 56.0\% & 50.0\%   \\
    \texttt{random}  & 44.0\% & 50.0\%  \\
    \midrule
    \texttt{vanilla} & 46.0\% & 48.0\%   \\
    \texttt{random}  & 54.0\% & 52.0\%  \\
    \bottomrule
  \end{tabular}
\end{wraptable}

\subsection{SHA-1 preimage attacks}

\begin{wrapfigure}{r}{200pt}
  \centering
  \includegraphics[width=200pt]{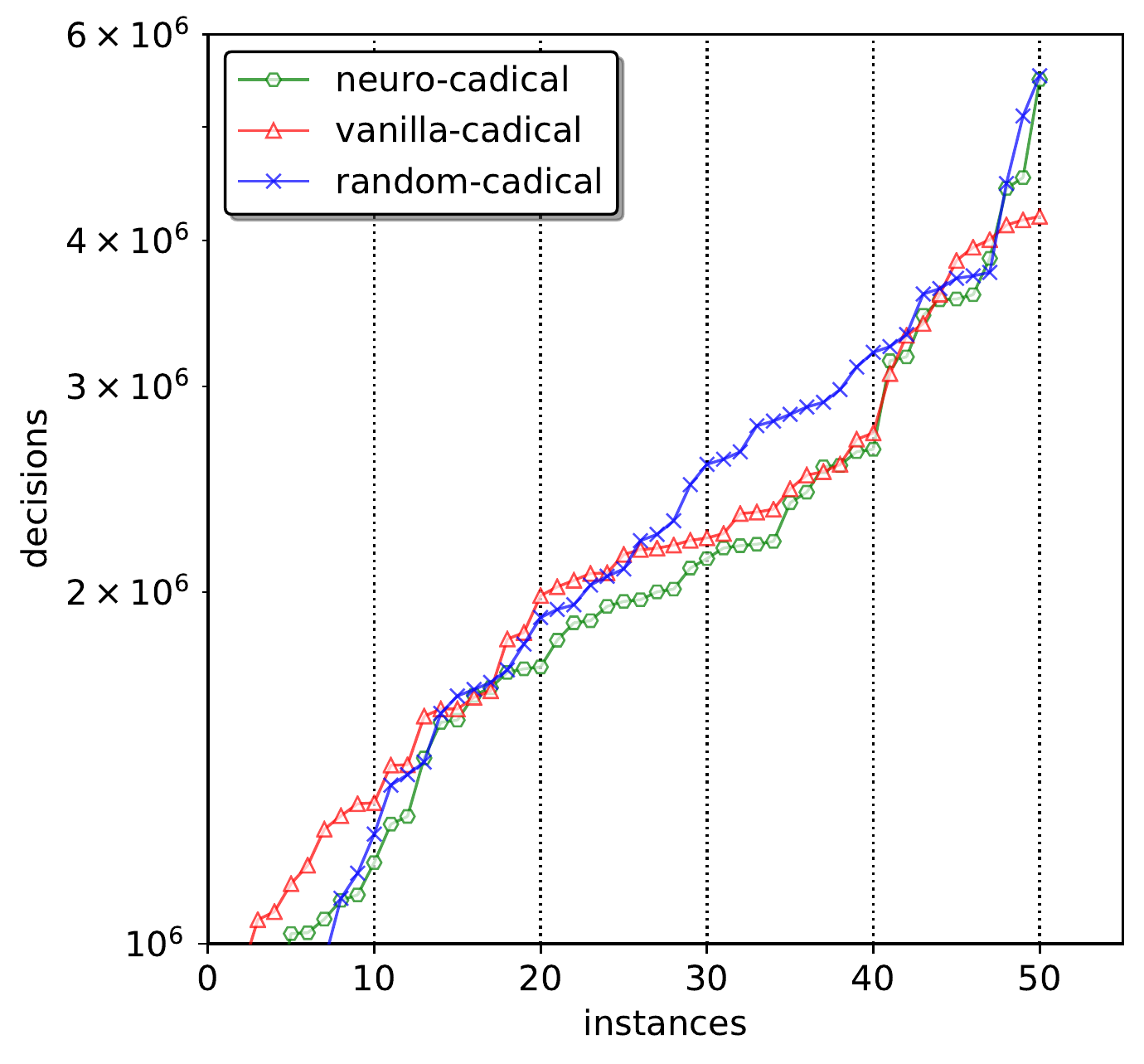}
    \caption{Decision cactus plot on the \texttt{sha-1} test dataset. For this evaluation, \texttt{neuro-cadical} was trained with reinforcement learning to minimize expected glue levels of conflict clauses. As before, \texttt{neuro-cadical} tends to be more decision efficient than either baseline.}   \label{fig:sha250-decisions}

  \end{wrapfigure}

In this section, \texttt{neuro-cadical} performs periodic refocusing with the version of NeuroGlue trained as a policy network via reinforcement learning (\Cref{subsect:reinforcement-learning}).
On our test dataset of \(50\) SHA-1 preimage attack problems, we perform \(7\) evaluation runs with distinct random seeds with the same hardware and a \(1000\) second timeout, averaging the results as in \Cref{subsect:sc-evals}. 
\texttt{neuro-cadical} achieves a PAR-2 score of \(279.29\), \(1.23\%\) better than that of \texttt{vanilla-cadical} (\(282.78\)) and \(8.19\%\) better than that of \texttt{random-cadical} (\(304.19\)). Figure \ref{fig:sha250-decisions} displays a decision cactus plot with 6M cutoff on the entire dataset. As in \Cref{subsect:sc-evals}, \texttt{neuro-cadical} tends to be more decision efficient than both baselines. From \Cref{tab:sha250-comp}, we see that, as in \Cref{subsect:sc-evals} \texttt{neuro-cadical} has better GLR on more problems than both baselines.

\section{Discussion}

\paragraph{Related work on glue variables} We note that
prioritizing the activity scores of variables related to glue clauses is not a novel idea, and dates back to Glucose, which additionally bumps the activity scores of variables in a learned clause which were propagated by a glue clause. The effectiveness of prioritizing glue variables in EVSIDS branching heuristics has already been demonstrated in \cite{DBLP:conf/cp/Chowdhury0Y19}, where \emph{glue bumping} was shown to improve solver performance on SAT competition benchmarks. However, in contrast to our approach, glue bumping is an online heuristic that only increases the score of variables which have \emph{already} frequently appeared in a glue clause for a single run of the solver, and does not attempt to \emph{predict} glue variables from the formula itself, as we do.

\paragraph{Future directions} Although in our present work, for the sake of simplicity, we have avoided exposing our policy network to any part of the solver state, the positive results of \cite{DBLP:conf/iclr/LedermanRSL20} indicate that even exposing very basic information can be beneficial for learning branching heuristics. We consider this to be a promising path to further improving solver performance.

\paragraph{Conclusion} We have proposed training for \emph{glue variable prediction} to guide SAT solvers through periodic refocusing, approaching the task in terms of both supervised and reinforcement learning. Along with a lightweight network architecture, we have demonstrated the effectiveness of both approaches by improving the performance of a state-of-the-art SAT solver on diverse benchmarks with no hardware acceleration, thus addressing the limitations of previous work in this vein \cite{DBLP:conf/sat/SelsamB19} and showing that we can realize the promise of neural networks for accelerating high-performance SAT solvers in all contexts in which they are currently deployed. We are optimistic that refinements to our approach, possibly incorporating solver state and history, will push the state-of-the-art even further.

\newpage

\begin{ack}
This work benefitted from conversations with John Harrison, Thomas Hales, and Daniel Selsam. We also thank Volodymyr Skladanivskyy for assistance with {\sc CGen}.
\end{ack}

\label{sect:bib}
\linespread{1}\selectfont
\bibliographystyle{plain}
\bibliography{neuro-cadical-neurips}

\begin{thebibliography}{10}

\bibitem{DBLP:journals/tcs/Alekhnovich04}
Michael Alekhnovich.
\newblock Mutilated chessboard problem is exponentially hard for resolution.
\newblock {\em Theor. Comput. Sci.}, 310(1-3):513--525, 2004.

\bibitem{DBLP:conf/iclr/AmizadehMW19}
Saeed Amizadeh, Sergiy Matusevych, and Markus Weimer.
\newblock Learning to solve circuit-sat: An unsupervised differentiable
  approach.
\newblock In {\em 7th International Conference on Learning Representations,
  {ICLR} 2019, New Orleans, LA, USA, May 6-9, 2019}. OpenReview.net, 2019.

\bibitem{DBLP:conf/ijcai/AudemardS09}
Gilles Audemard and Laurent Simon.
\newblock Predicting learnt clauses quality in modern {SAT} solvers.
\newblock In Craig Boutilier, editor, {\em {IJCAI} 2009, Proceedings of the
  21st International Joint Conference on Artificial Intelligence, Pasadena,
  California, USA, July 11-17, 2009}, pages 399--404, 2009.

\bibitem{DBLP:journals/ijait/AudemardS18}
Gilles Audemard and Laurent Simon.
\newblock On the glucose {SAT} solver.
\newblock {\em Int. J. Artif. Intell. Tools}, 27(1):1840001:1--1840001:25,
  2018.

\bibitem{DBLP:journals/corr/BaKH16}
Lei~Jimmy Ba, Jamie~Ryan Kiros, and Geoffrey~E. Hinton.
\newblock Layer normalization.
\newblock {\em CoRR}, abs/1607.06450, 2016.

\bibitem{DBLP:conf/icml/BansalLRSW19}
Kshitij Bansal, Sarah~M. Loos, Markus~N. Rabe, Christian Szegedy, and Stewart
  Wilcox.
\newblock Holist: An environment for machine learning of higher order logic
  theorem proving.
\newblock In Kamalika Chaudhuri and Ruslan Salakhutdinov, editors, {\em
  Proceedings of the 36th International Conference on Machine Learning, ICML
  2019, 9-15 June 2019, Long Beach, California, {USA}}, volume~97 of {\em
  Proceedings of Machine Learning Research}, pages 454--463. {PMLR}, 2019.

\bibitem{DBLP:journals/apal/BeameP96}
Paul Beame and Toniann Pitassi.
\newblock An exponential separation between the parity principle and the
  pigeonhole principle.
\newblock {\em Ann. Pure Appl. Logic}, 80(3):195--228, 1996.

\bibitem{biere2017cadical}
Armin Biere.
\newblock {CaDiCaL}, {L}ingeling, {P}lingeling, {T}reengeling and {YalSAT}
  entering the {SAT} {C}ompetition 2017.
\newblock {\em Proc. of SAT Competition 2017}, pages 14--15, 2017.

\bibitem{biere2018cadical}
Armin Biere.
\newblock {CaDiCaL}, {L}ingeling, {P}lingeling, {T}reengeling and {YalSAT}
  entering the {SAT} {C}ompetition 2018.
\newblock {\em Proc. of SAT Competition 2018}, pages 13--14, 2018.

\bibitem{biere2019cadical}
Armin Biere.
\newblock {CaDiCaL} at the {SAT} {R}ace 2019.
\newblock {\em Proceedings of SAT Race (2019, Submitted)}, 2019.

\bibitem{DBLP:conf/issac/BrainDG17}
Martin Brain, James~H. Davenport, and Alberto Griggio.
\newblock Benchmarking solvers, sat-style.
\newblock In Matthew England and Vijay Ganesh, editors, {\em Proceedings of the
  2nd International Workshop on Satisfiability Checking and Symbolic
  Computation co-located with the 42nd International Symposium on Symbolic and
  Algebraic Computation {(ISSAC} 2017), Kaiserslautern, Germany, July 29,
  2017}, volume 1974 of {\em {CEUR} Workshop Proceedings}. CEUR-WS.org, 2017.

\bibitem{DBLP:conf/cp/Chowdhury0Y19}
Md.~Solimul Chowdhury, Martin M{\"{u}}ller, and Jia{-}Huai You.
\newblock Exploiting glue clauses to design effective {CDCL} branching
  heuristics.
\newblock In Thomas Schiex and Simon de~Givry, editors, {\em Principles and
  Practice of Constraint Programming - 25th International Conference, {CP}
  2019, Stamford, CT, USA, September 30 - October 4, 2019, Proceedings}, volume
  11802 of {\em Lecture Notes in Computer Science}, pages 126--143. Springer,
  2019.

\bibitem{DBLP:conf/stoc/Cook71}
Stephen~A. Cook.
\newblock The complexity of theorem-proving procedures.
\newblock In Michael~A. Harrison, Ranan~B. Banerji, and Jeffrey~D. Ullman,
  editors, {\em Proceedings of the 3rd Annual {ACM} Symposium on Theory of
  Computing, May 3-5, 1971, Shaker Heights, Ohio, {USA}}, pages 151--158.
  {ACM}, 1971.

\bibitem{cook1976short}
Stephen~A Cook.
\newblock A short proof of the pigeon hole principle using extended resolution.
\newblock {\em Acm Sigact News}, 8(4):28--32, 1976.

\bibitem{cook1979relative}
Stephen~A Cook and Robert~A Reckhow.
\newblock The relative efficiency of propositional proof systems.
\newblock {\em The Journal of Symbolic Logic}, 44(1):36--50, 1979.

\bibitem{DBLP:journals/cacm/DavisLL62}
Martin Davis, George Logemann, and Donald~W. Loveland.
\newblock A machine program for theorem-proving.
\newblock {\em Commun. {ACM}}, 5(7):394--397, 1962.

\bibitem{DBLP:journals/jacm/DavisP60}
Martin Davis and Hilary Putnam.
\newblock A computing procedure for quantification theory.
\newblock {\em J. {ACM}}, 7(3):201--215, 1960.

\bibitem{neural-heuristics-for-sat}
Sebastian Jaszczur, Micha{\l} {\L}uszczyk, and Henryk Michalewski.
\newblock Neural heuristics for {SAT} solving.
\newblock In {\em Representation Learning on Graphs and Manifolds Workshop at
  ICLR 2019}, 2019.

\bibitem{kurin2019improving}
Vitaly Kurin, Saad Godil, Shimon Whiteson, and Bryan Catanzaro.
\newblock Improving {SAT} solver heuristics with graph networks and
  reinforcement learning.
\newblock {\em arXiv preprint arXiv:1909.11830}, 2019.

\bibitem{DBLP:conf/iclr/LedermanRSL20}
Gil Lederman, Markus~N. Rabe, Sanjit Seshia, and Edward~A. Lee.
\newblock Learning heuristics for quantified boolean formulas through
  reinforcement learning.
\newblock In {\em 8th International Conference on Learning Representations,
  {ICLR} 2020, Addis Ababa, Ethiopia, April 26-30, 2020}. OpenReview.net, 2020.

\bibitem{DBLP:conf/ijcai/LiangKPCG18}
Jia Liang, Hari Govind~V. K., Pascal Poupart, Krzysztof Czarnecki, and Vijay
  Ganesh.
\newblock An empirical study of branching heuristics through the lens of global
  learning rate.
\newblock In J{\'{e}}r{\^{o}}me Lang, editor, {\em Proceedings of the
  Twenty-Seventh International Joint Conference on Artificial Intelligence,
  {IJCAI} 2018, July 13-19, 2018, Stockholm, Sweden}, pages 5319--5323.
  ijcai.org, 2018.

\bibitem{DBLP:phd/basesearch/Liang18a}
Jia~Hui Liang.
\newblock {\em Machine Learning for {SAT} Solvers}.
\newblock PhD thesis, University of Waterloo, Ontario, Canada, 2018.

\bibitem{DBLP:conf/sat/LiangGPC16}
Jia~Hui Liang, Vijay Ganesh, Pascal Poupart, and Krzysztof Czarnecki.
\newblock Learning rate based branching heuristic for {SAT} solvers.
\newblock In Nadia Creignou and Daniel~Le Berre, editors, {\em Theory and
  Applications of Satisfiability Testing - {SAT} 2016 - 19th International
  Conference, Bordeaux, France, July 5-8, 2016, Proceedings}, volume 9710 of
  {\em Lecture Notes in Computer Science}, pages 123--140. Springer, 2016.

\bibitem{DBLP:conf/lpar/LoosISK17}
Sarah~M. Loos, Geoffrey Irving, Christian Szegedy, and Cezary Kaliszyk.
\newblock Deep network guided proof search.
\newblock In Thomas Eiter and David Sands, editors, {\em LPAR-21, 21st
  International Conference on Logic for Programming, Artificial Intelligence
  and Reasoning, Maun, Botswana, May 7-12, 2017}, volume~46 of {\em EPiC Series
  in Computing}, pages 85--105. EasyChair, 2017.

\bibitem{DBLP:conf/osdi/MoritzNWTLLEYPJ18}
Philipp Moritz, Robert Nishihara, Stephanie Wang, Alexey Tumanov, Richard Liaw,
  Eric Liang, Melih Elibol, Zongheng Yang, William Paul, Michael~I. Jordan, and
  Ion Stoica.
\newblock Ray: {A} distributed framework for emerging {AI} applications.
\newblock In Andrea~C. Arpaci{-}Dusseau and Geoff Voelker, editors, {\em 13th
  {USENIX} Symposium on Operating Systems Design and Implementation, {OSDI}
  2018, Carlsbad, CA, USA, October 8-10, 2018}, pages 561--577. {USENIX}
  Association, 2018.

\bibitem{DBLP:conf/dac/MoskewiczMZZM01}
Matthew~W. Moskewicz, Conor~F. Madigan, Ying Zhao, Lintao Zhang, and Sharad
  Malik.
\newblock Chaff: Engineering an efficient {SAT} solver.
\newblock In {\em Proceedings of the 38th Design Automation Conference, {DAC}
  2001, Las Vegas, NV, USA, June 18-22, 2001}, pages 530--535, 2001.

\bibitem{DBLP:conf/sat/Oh15}
Chanseok Oh.
\newblock Between {SAT} and {UNSAT:} the fundamental difference in {CDCL}
  {SAT}.
\newblock In Marijn Heule and Sean~A. Weaver, editors, {\em Theory and
  Applications of Satisfiability Testing - {SAT} 2015 - 18th International
  Conference, Austin, TX, USA, September 24-27, 2015, Proceedings}, volume 9340
  of {\em Lecture Notes in Computer Science}, pages 307--323. Springer, 2015.

\bibitem{DBLP:conf/nips/PaszkeGMLBCKLGA19}
Adam Paszke, Sam Gross, Francisco Massa, Adam Lerer, James Bradbury, Gregory
  Chanan, Trevor Killeen, Zeming Lin, Natalia Gimelshein, Luca Antiga, Alban
  Desmaison, Andreas K{\"{o}}pf, Edward Yang, Zachary DeVito, Martin Raison,
  Alykhan Tejani, Sasank Chilamkurthy, Benoit Steiner, Lu~Fang, Junjie Bai, and
  Soumith Chintala.
\newblock {PyTorch}: An imperative style, high-performance deep learning
  library.
\newblock In Hanna~M. Wallach, Hugo Larochelle, Alina Beygelzimer, Florence
  d'Alch{\'{e}}{-}Buc, Emily~B. Fox, and Roman Garnett, editors, {\em Advances
  in Neural Information Processing Systems 32: Annual Conference on Neural
  Information Processing Systems 2019, NeurIPS 2019, 8-14 December 2019,
  Vancouver, BC, Canada}, pages 8024--8035, 2019.

\bibitem{10.1137/0330046}
B.~T. Polyak and A.~B. Juditsky.
\newblock Acceleration of stochastic approximation by averaging.
\newblock {\em SIAM J. Control Optim.}, 30(4):838–855, July 1992.

\bibitem{satcomp-website}
{S}atisfiability: {A}pplication and {T}heory ({SAT})~e.{V}.
\newblock The {I}nternational {SAT} {c}ompetition web page.
\newblock https://http://www.satcompetition.org/.

\bibitem{DBLP:journals/tnn/ScarselliGTHM09}
Franco Scarselli, Marco Gori, Ah~Chung Tsoi, Markus Hagenbuchner, and Gabriele
  Monfardini.
\newblock The graph neural network model.
\newblock {\em {IEEE} Trans. Neural Networks}, 20(1):61--80, 2009.

\bibitem{DBLP:conf/sat/SelsamB19}
Daniel Selsam and Nikolaj Bj{\o}rner.
\newblock Guiding high-performance {SAT} solvers with unsat-core predictions.
\newblock In {\em Theory and Applications of Satisfiability Testing - {SAT}
  2019 - 22nd International Conference, {SAT} 2019, Lisbon, Portugal, July
  9-12, 2019, Proceedings}, pages 336--353, 2019.

\bibitem{DBLP:journals/corr/abs-1903-04671}
Daniel Selsam and Nikolaj Bj{\o}rner.
\newblock Neurocore: Guiding high-performance {SAT} solvers with unsat-core
  predictions.
\newblock {\em CoRR}, abs/1903.04671, 2019.

\bibitem{DBLP:conf/iclr/SelsamLBLMD19}
Daniel Selsam, Matthew Lamm, Benedikt B{\"{u}}nz, Percy Liang, Leonardo
  de~Moura, and David~L. Dill.
\newblock Learning a {SAT} solver from single-bit supervision.
\newblock In {\em 7th International Conference on Learning Representations,
  {ICLR} 2019, New Orleans, LA, USA, May 6-9, 2019}, 2019.

\bibitem{DBLP:conf/ictai/SilvaS96}
Jo{\~{a}}o P.~Marques Silva and Karem~A. Sakallah.
\newblock Conflict analysis in search algorithms for satisfiability.
\newblock In {\em Eighth International Conference on Tools with Artificial
  Intelligence, ICTAI '96, Toulouse, France, November 16-19, 1996}, pages
  467--469. {IEEE} Computer Society, 1996.

\bibitem{cgen}
Volodymyr Skladanivskyy.
\newblock Cgen: a tool for encoding sha-1 and sha-256 hash functions into cnf
  in dimacs format.

\bibitem{vsklad2}
Volodymyr Skladanivskyy.
\newblock Minimalistic round-reduced {SHA}-1 pre-image attack.
\newblock {\em Proceedings of SAT Race (2019, Submitted)}, 2019.

\bibitem{vsklad1}
Volodymyr Skladanivskyy.
\newblock Tailored compact {CNF} encodings for {SHA}-1.
\newblock {\em J. Satisf. Boolean Model. Comput.}, 2020.
\newblock Under review.

\bibitem{torchscript}
The~{PyTorch} team.
\newblock {T}orch {S}cript.
\newblock https://pytorch.org/docs/stable/jit.html.

\bibitem{tseitin1983complexity}
Grigori~S Tseitin.
\newblock On the complexity of derivation in propositional calculus.
\newblock In {\em Automation of reasoning}, pages 466--483. Springer, 1983.

\bibitem{DBLP:journals/corr/abs-1904-12084}
Zhanfu Yang, Fei Wang, Ziliang Chen, Guannan Wei, and Tiark Rompf.
\newblock Graph neural reasoning for 2-quantified boolean formula solvers.
\newblock {\em CoRR}, abs/1904.12084, 2019.

\bibitem{DBLP:conf/nips/YolcuP19}
Emre Yolcu and Barnab{\'{a}}s P{\'{o}}czos.
\newblock Learning local search heuristics for boolean satisfiability.
\newblock In Hanna~M. Wallach, Hugo Larochelle, Alina Beygelzimer, Florence
  d'Alch{\'{e}}{-}Buc, Emily~B. Fox, and Roman Garnett, editors, {\em Advances
  in Neural Information Processing Systems 32: Annual Conference on Neural
  Information Processing Systems 2019, NeurIPS 2019, 8-14 December 2019,
  Vancouver, BC, Canada}, pages 7990--8001, 2019.

\end{thebibliography}

\end{document}